\documentclass[12pt]{article}
\usepackage{pdproc,graphicx} 

\begin{document}

\renewcommand{\thefootnote}{\alph{footnote}}
  
\title{
NEUTRINO OSCILLATIONS, GLOBAL ANALYSIS AND THETA(13)}

\author{G.L.~FOGLI~$^{1,2*}$, E.~LISI~$^{2}$, A.~MARRONE~$^{1,2}$, 
A.~PALAZZO~$^3$,
A.M.~ROTUNNO~$^{1,2}$}
\smallskip
\address{ $^1$~Dipartimento di Fisica, Universit\`a di Bari,
Via Amendola n.176, 70126 Bari, Italy \\[+0.9mm]
$^2$~Sezione INFN di Bari, Via Orabona n.4, 70126 Bari, Italy\\[+0.4mm]
$^3$~AHEP Group, Institut de F\'isica Corpuscular, CSIC/Universitat de Val\`encia,\\
Edifici Instituts d'Investigaci\'o, Apt.\ 22085, 46071 Val\`encia, Spain }

\centerline{\footnotesize $^*$Speaker. \tt fogli@ba.infn.it}
\abstract{At the previous Venice meeting NO-VE 2008, we discussed
possible hints in favor of a nonzero value for the
unknown neutrino mixing angle $\theta_{13}$, emerging from the combination
of solar and long-baseline reactor data, as well as from the combination of
atmospheric, CHOOZ and long-baseline accelerator $\nu_\mu\to\nu_\mu$ data. 
Recent  MINOS 2009 results in the
$\nu_\mu\to\nu_e$ appearance channel also seem to support such hints. 
A combination of all current oscillation data provides, as preferred range,
$$
\sin^2\theta_{13}\simeq 0.02\pm 0.01 \ \ (1\sigma)\ .
$$
We review several issues raised by such hints in the last year, 
and comment on their possible near-future improvements and  tests.}

\normalsize\baselineskip=15pt


\section{Introduction}

Neutrino flavor oscillations induced by  $\nu$ masses and mixings 
are by now well established. The $3\nu$ oscillation parameter space includes
two independent mass-squared differences
($\delta m^2,\,\Delta m^2$), and three 
mixing angles $(\theta_{12},\,\theta_{23},\,\theta_{13})$; see \cite{Foglireview} 
for a review. Among these parameters, only $\theta_{13}$ is compatible with
zero, with an upper bound dominated by the CHOOZ experiment \cite{CHOOZ} in the range
$\sin^2\theta_{13}<$~few~\%. 

A nonzero value of $\theta_{13}$ is crucial to access
a possible CP-violating phase $\delta$ \cite{Bernabeu,Maltoni}, 
to probe the $\nu$ mass hierarchy [i.e., sign($\pm\Delta m^2$)] 
via $\nu$ interactions in the Earth \cite{Maltoni,Cowen}
or in Supernovae \cite{Yuksel,Lisi}, and to discriminate different theoretical models for the neutrino mass matrix
\cite{Albright,Scott,King,Altarelli}. Therefore, the search for $\theta_{13}$ is a primary objective
in the $\nu$ physics program worldwide \cite{Spiering,Katsanevas,Lande,Diwan,Mezzetto,Kajita,Cao,Dawson}, and
its determination will also be important for longer-term goals related to the search for
new neutrino interactions \cite{Meloni,Minakata,Lindner} or new physics at very high scales relevant 
for leptogenesis \cite{Petcov}.

Given the large and growing interest in the value of $\theta_{13}$, 
it is worthwhile to investigate if the available data can provide
at least some ``hints'' in favor of $\theta_{13}>0$, at the level of $\sim\!\!1$--2$\sigma$, before 
an evidence at $>\!3\sigma$ is hopefully found in
dedicated searches at reactors \cite{Cao,Dawson} and accelerators \cite{Mezzetto,Kajita}.  

\newpage

\section{First hints of $\theta_{13}>0$ (2008)}

\vspace*{-4mm}
\subsection{Solar+KamLAND neutrino data}

At the last Workshop ``Neutrino Oscillations in Venice'' (NO-VE 2008), we noted~\cite{NOVE2008}
that the combination of solar data and long-baseline
reactor data (KamLAND \cite{Suzuki}) suggested a weak preference for $\sin^2\theta_{13}\sim O(10^{-2})$ at $\sim\!\!0.5\sigma$,
as a result of a slight difference between the best-fit values of $\sin^2\theta_{12}$ in the
two datasets. 

Shortly after, the same effect was independently discussed in Ref.~\cite{Bala}.
One month later, at the Neutrino~2008 Conference \cite{Nu2008}, new solar data were presented 
from the third phase of the Sudbury Neutrino Observatory (SNO-III) \cite{SNO2008,McDonald}. 
We then showed that such data corroborated the previous picture at the level of $\sim\!\! 1.2\sigma$ 
\cite{theta13}
\begin{equation}
\label{hint1}
\sin^2\theta_{13}\simeq 0.021\pm 0.017\ (1\sigma, \ \mathrm{solar}+\mathrm{KamLAND},\ 2008)\ .
\end{equation}
A similar hint (at a sightly higher confidence level of $\sim\!\! 1.5\sigma$) was obtained in 
Ref.~\cite{Valle}, and is also implicit in the results of Ref.~\cite{Pena}.
So, several
independent analyses have found a weak 
 preference in favor of $\theta_{13}>0$,
using the available (2008) solar and KamLAND data, at the level of 1.2--1.5$\sigma$. 

\begin{figure}[t]
\vspace*{0cm}
\hspace*{.1cm}
\includegraphics[height=7.7cm,width=14cm]{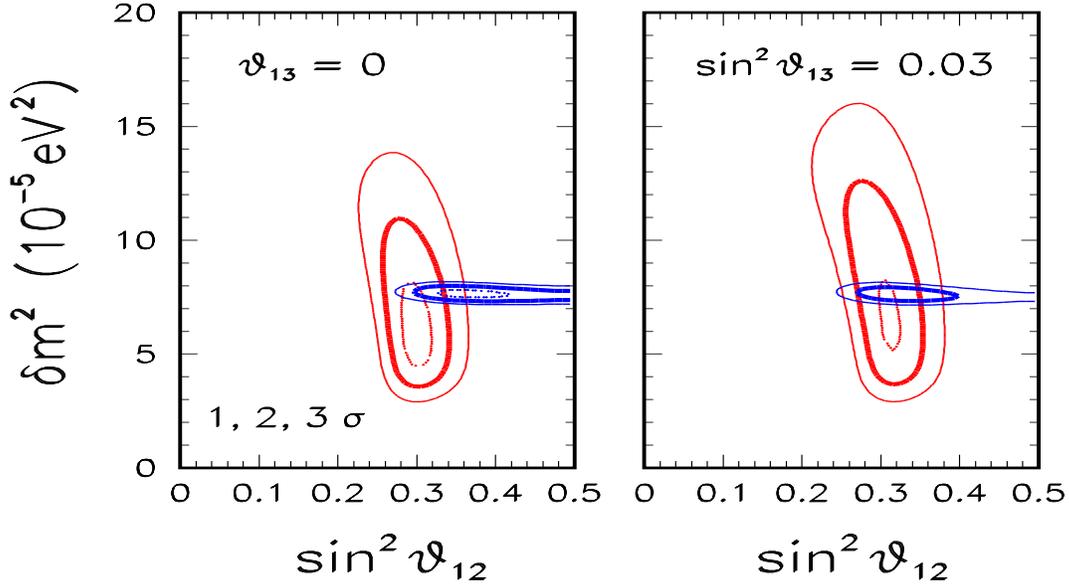}
\vspace*{0mm}
\caption{Comparison of $n$-$\sigma$ regions allowed by the latest (2008)
solar and KamLAND data in the $(\delta m^2,\sin^2\theta_{12})$ plane, 
for two fixed values of $\theta_{13}$. \vspace*{-10mm}}
\end{figure}

Figure~1 shows the effect of nonzero $\theta_{13}$ on the regions 
separately allowed by the latest available data from the
 solar  and KamLAND experiments, at 1, 2 and $3\sigma$ level
(i.e., $\Delta\chi^2=1$, 4 and 9, at fixed $\theta_{13}$). The left and right panels
refer to $\theta_{13}=0$ and to a representative value $\sin^2\theta_{13}=0.03$, respectively. Clearly,  
the two best-fit regions overlap more for $\sin^2\theta_{13}>0$.

\begin{figure}[t]
\hspace{1pc}%
\begin{minipage}{17.5pc}
\includegraphics[width=7cm]{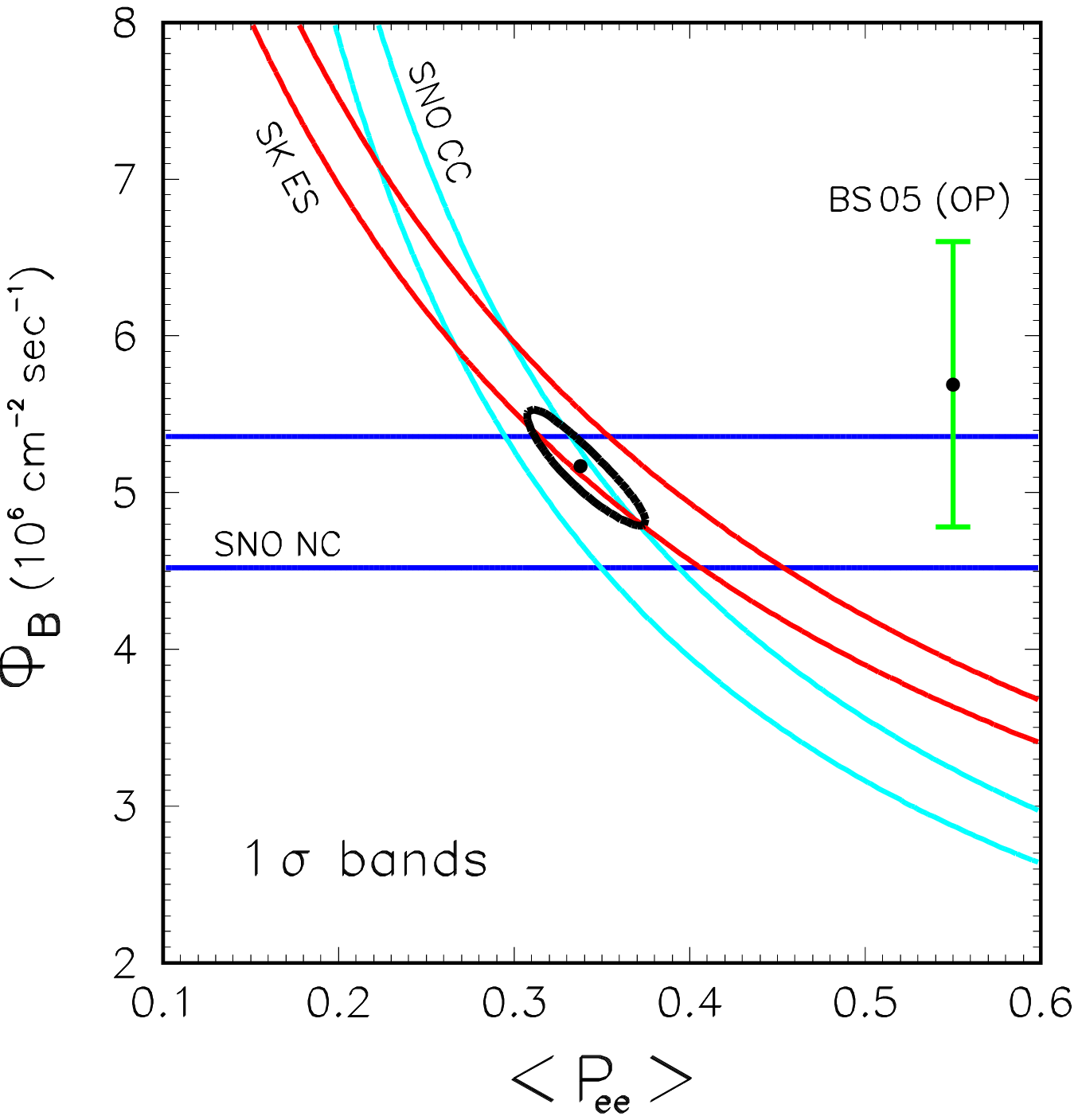}
\end{minipage}
\hspace{1pc}%
\begin{minipage}{17.5pc}
\hspace{-1pc}
\includegraphics[width=7cm]{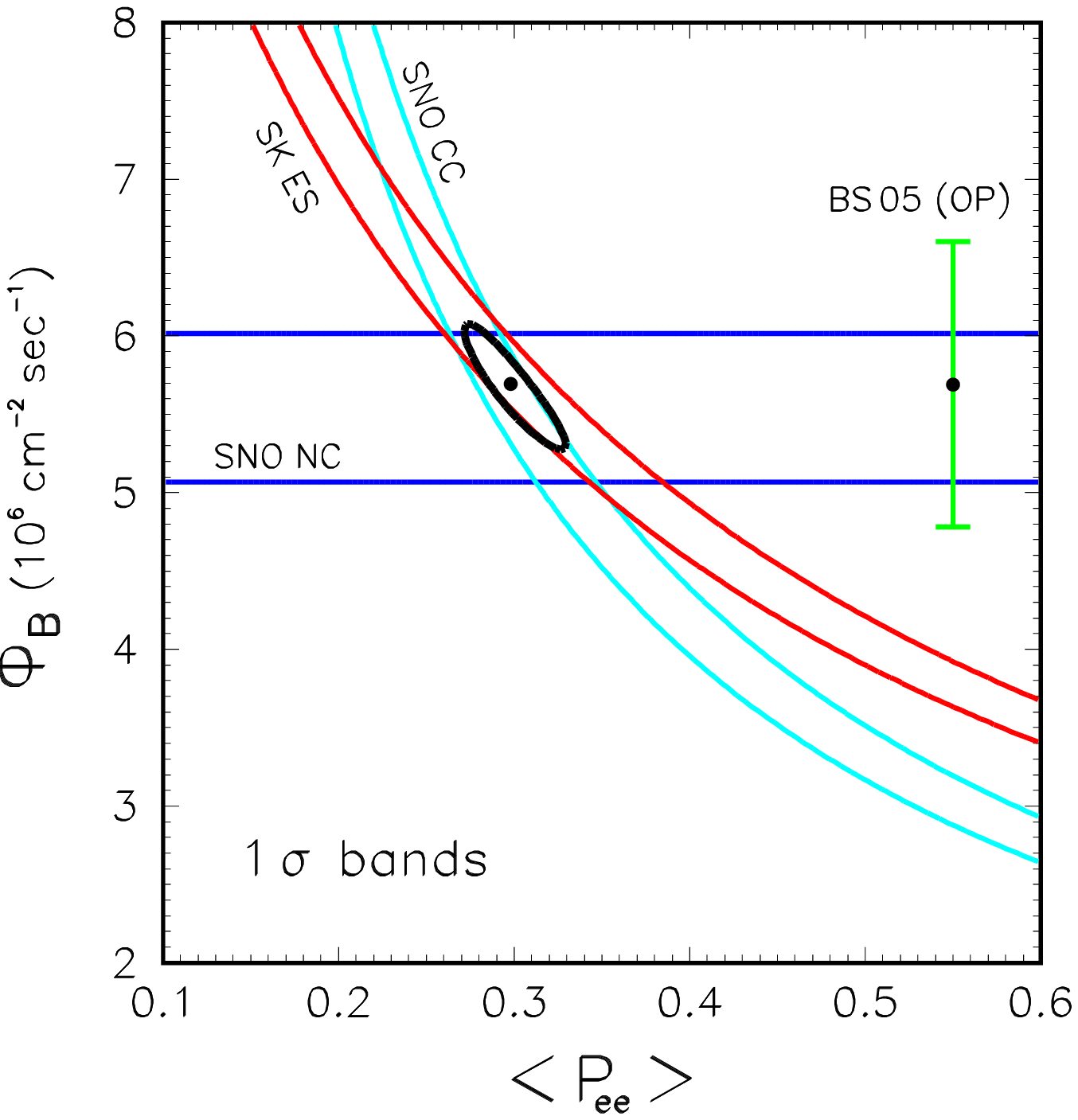}
\end{minipage} 
\vspace*{2mm}
\caption{Model-independent analysis of $^8$B solar $\nu$ data in SK and SNO,
in the plane charted by the $\Phi_\mathrm{B}$ flux and by the energy-averaged survival probability $P_{ee}$.
The bands represent the regions allowed at $1\sigma$ by the NC and CC event rates measured in SNO and by
the ES rate measured in SK. The left panel includes only SNO-I and SNO-II data as available in 2005,
while the right panel includes only SNO-III data as available in 2008.
In each panel, the slanted ellipse represents the combination of NC+CC+ES constraints, while
the vertical error bar represents the $\Phi_\mathrm{B}$ range predicted by the BS'05 (OP) standard solar model.
\vspace*{5mm}}
\end{figure}

Notably, there is no reason to think
that the preference for $\theta_{13}>0$ may be due to a 
``fluctuation'' of the
latest SNO-III data, which are actually
 in very good agreement with previous SNO and Super-Kamiokande (SK) solar $\nu$
data, as we now show. 

Figure~2 updates earlier 
model-independent (and also $\theta_{13}$-independent) analyses~\cite{Modelind,Getting,Foglireview} 
of Super-Kamiokande elastic scattering (SK ES) data \cite{Naka} and SNO neutral-current (NC)
and charged-current (CC) 
data \cite{McDonald}, 
in the plane charted by the $^8$B $\nu_e$ flux $(\Phi_\mathrm{B})$ and by
the corresponding, energy-averaged survival probability $\langle P_{ee} \rangle$.
The left panel refers to SNO 2005 data (i.e., from SNO phase I and II only), 
while the right panel includes only SNO-III data as available in 2008. The 
relative agreement among the $1\sigma$ bands constrained by the ES, NC, and CC data,
which is already good in the left panel, becomes even better in the right
panel. Therefore, the SNO-III data 
do improve the overall consistency of the high-statistics (ES, NC, CC) data 
 collected in SK and SNO.%
\footnote{In the right plot of Fig.~2, we also note that 
the global SK+SNO combination (slanted ellipse) compares very well with with the $^8$B flux
predictions of a reference ``standard solar model'' (SSM Bahcall-Serenelli-Basu 2005 OP) \cite{BS05}. However, 
unsolved metallicity  discrepancies \protect\cite{Pena} still 
prevent a more quantitative comparison of $^8$B neutrino data with SSM's.}
\newpage

\subsection{Super-Kamiokande atmospheric neutrino data}
\vspace*{-1.5mm}

Besides the solar+KamLAND ``hint,'' which is relatively recent, 
an older, independent preference for $\theta_{13}>0$ 
had already been found in \cite{Foglireview}
(at the level of $\sim\!\!0.9\sigma$), from an analysis of atmospheric 
neutrino data in phase-I of Super-Kamiokande (SK-I), together with CHOOZ and long-baseline accelerator
(LBL) data.
We traced its origin to
subleading $3\nu$ oscillation terms, arising at first order in $\theta_{13}$ and
driven by $\delta m^2$ \cite{Peres}, which are 
most effective at $\cos\delta=-1$; see, in particular, Fig.~24 in \cite{Foglireview}.
The related effects  
could partly explain the observed excess of sub-GeV atmospheric
electron-like events. This weak hint for $\theta_{13}>0$ is compatible with
the CHOOZ constraints, and is not spoiled by adding LBL data
from K2K and MINOS in the
$\nu_\mu$ disappearance channel, which are not yet sensitive to $\theta_{13}$ \cite{Melch}. Our
constrained estimate reads \cite{theta13}
\vspace*{-2mm}
\begin{equation}
\label{hint2}
\sin^2\theta_{13}\simeq 0.012\pm 0.013\ (1\sigma, \ \mathrm{Atmos.}+\mathrm{LBL}+\mathrm{CHOOZ},\ 2008)\ .
\end{equation}
\vspace*{-0.5mm}
Unfortunately, subleading ($\delta m^2,\,\theta_{13})$ effects  are  generally
small in the relevant low-energy
atmospheric $\nu$ distributions, and are
smeared  by the interaction and detection processes \cite{Foglireview,Maltoni}. Therefore,
their emergence may depend on details of the statistical analysis. An independent,
state-of-the-art $3\nu$  global fit of atmospheric $\nu$ data did not find an appreciable 
preference for $\theta_{13}>0$ \protect\cite{Concha}, while another (less documented) 
analysis \protect\cite{Ernst1}  appeared to favor $-\!\cos\delta\sin \theta_{13}>0$ as in our case;
see also Ref.\cite{Ernst2}.
\footnote{In \protect\cite{Ernst1,Ernst2},
the appearance of ``negative'' values of  $\theta_{13}$ is the result of an
unusual convention, which absorbs into $\theta_{13}$ the negative sign associated to
the case $\cos\delta <0$.}

The authors of Ref.~\cite{Schwetz} analyzed several variants in the atmos.+LBL+CHOOZ fit, and found
either no hint or, at most, a $0.5\sigma$ one. The latter, although 
weaker than our $0.9\sigma$ [Eq.~(2)], shows similar qualitative features, such
as the role of $\delta m^2$-driven terms and the irrelevance of current
LBL $\nu_\mu$ disappearance data. It was also observed \cite{Schwetz} that the (still unpublished) SK-II data 
\cite{Naka}
prefer $\theta_{13}\simeq 0$, due to a small deficit (rather than an excess) of upgoing 
multi-GeV electron-like (MG$e$) events. However, we note
that preliminary SK-III data seem to show again a small excess in the same event
sample \cite{Nu2008}. The full
analysis of SK-II (and possibly of SK-III and even SK-IV) atmospheric data
is underway \cite{Naka} and is likely to affect the $\theta_{13}$ constraints.

Unfortunately, the number of SK bins is growing to $O(10^3)$, with $O(10^2)$ 
systematics~\cite{Naka,Takenaga,Wendell}, making it impossible to closely reproduce their 
data analysis.
It is then crucial that the SK Collaboration itself performs
a full $3\nu$ analysis including all ($\delta m^2,\theta_{13}$) 
oscillation terms \cite{Workshop}. Recent PhD 
thesis works in SK have assumed either $\delta m^2>0$ but with $\theta_{13}=0$ (see Ref.~\cite{Takenaga}) 
or  $\theta_{13}>0$ but with $\delta m^2=0$ (see Ref.~\cite{Wendell}). The latter approximation 
is also used in \cite{Valle}.
The $3\nu$ approximations adopted in Refs.~\cite{Takenaga,Wendell,Valle} 
prevent, at present, a direct comparison of their atmospheric $\nu$ results
with ours, as far as subtle effects driven by subleading ($\delta m^2,\,\theta_{13})$ terms  are concerned.

\begin{figure}[t]
\vspace*{0cm}
\hspace*{1.2cm}
\includegraphics[height=12.cm]{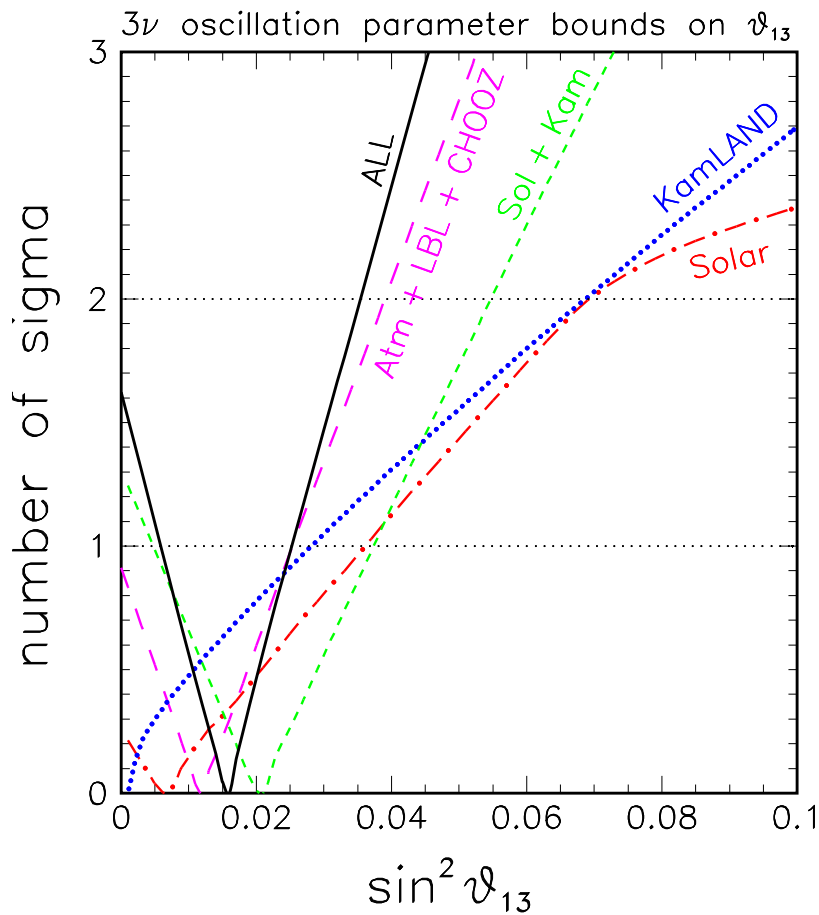}
\vspace*{0mm}
\caption{Bounds on $\theta_{13}$ from different data sets available in 2008. \vspace*{-10mm}}
\end{figure}
\vspace*{-9mm}

\subsection{Combination of all oscillation data (2008)}

We have presented two hints in favor of $\theta_{13}>0$, one coming from the analysis
of solar+KamLAND data [Eq.(1)], and another one from the analysis of published SK-I atmospheric
data, together with disappearance constraints from CHOOZ $\nu_e$ and LBL $\nu_\mu$ data
[Eq.~(2)]. As previously discussed, the second hint appears to have a more fragile status than the first, but
we have no compelling reason to revise it at present. Then,
by merging the results in Eqs.~(1) and (2) in a global neutrino data analysis, 
we obtain \cite{theta13}: 
\begin{equation}
\sin^2\theta_{13}\simeq 0.016\pm 0.010\ (1\sigma,\ \mathrm{All\ Data,\ 2008})\ ,
\end{equation}
which represents an intriguing indication  in favor of $\theta_{13}>0$ at the 90\% C.L.\ 
($\sim\!\!1.6\sigma)$.

Figure~3 summarizes our findings, by showing
the $n$-$\sigma$ curves ($n$-$\sigma=\sqrt{\Delta\chi^2}$) as a function of $\sin^2\theta_{13}$ 
(all other oscillation parameters being marginalized) for different combinations
of data sets available in 2008. The global combination (thick solid curve) provides, at $1\sigma$, the
range reported in Eq.~(3).

\newpage

\begin{figure}[t]
\vspace*{0cm}
\hspace*{0.5cm}
\includegraphics[height=5.cm]{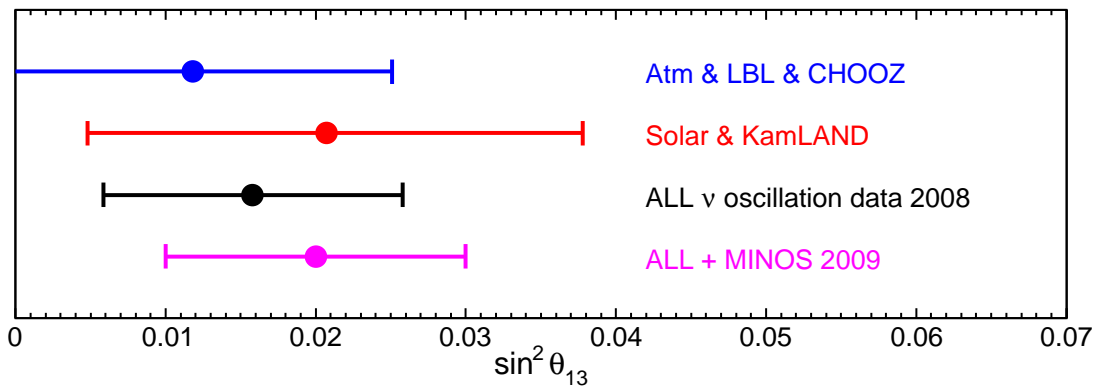}
\vspace*{0mm}
\caption{Hints of $\theta_{13}>0$ from different data sets and combinations: $1\sigma$ ranges. \vspace*{-10mm}}
\end{figure}
\vspace*{-2mm}

\section{MINOS $\nu_e$ appearance results and $\theta_{13}$ update (2009)}

At this Workshop, recent preliminary MINOS data 
in the $\nu_\mu\to\nu_e$ appearance channel have been presented  \cite{Diwan}. The data show
a slight overall excess of electron flavor events  
above the estimated background ($\Delta N_e \simeq 8\pm 5$), with a statistical significance of $\sim\! 90\%$~C.L.

If the overall excess $\Delta N_e$ is interpreted in terms of neutrino oscillations, 
and if the dominant oscillation parameters $(\Delta m^2,\,\sin^2\theta_{23})$ 
are fixed at their best-fit values, then one can put degenerate constraints
in the plane of the subdominant parameters $(\theta_{13},\,\delta)$ \cite{Diwan}.
In particular, the degenerate best-fit range is 
$\sin^2\theta_{13}\simeq 0.03$--0.05, depending on $\delta$ (and on the hierarchy),
while the 90\% C.L. bounds are at $\sin^2\theta_{13}\simeq 0$ (lower) and at 
$\sin^2\theta_{13}\simeq 0.07$--0.12 (upper). 
We tentatively symmetrize the preliminary MINOS 2009 preferred range with just
one significant digit as: 
\begin{equation}
\label{hint3}
\sin^2\theta_{13}\simeq 0.05\pm 0.03\ (1\sigma,\ \mathrm{MINOS}\ 2009)\ ,
\end{equation}
second-digit details being unimportant in this context.%
\footnote{MINOS $\nu_e$ appearance
results  can be properly included only after detailed publications of data and of their analyses become available.}

Thus, we might have two independent hints of $\theta_{13}>0$ at $90\%$ C.L.: one
from the global analysis of 2008 data [Eq.(3)], and one from the preliminary MINOS data
[Eq.~(4)]. Their combination at face value reads: 
\begin{equation}
\label{hint4}
\sin^2\theta_{13}\simeq 0.02\pm 0.01\ (1\sigma,\ \mathrm{All\ data}+\mathrm{MINOS}~2009)\ ,
\end{equation}
which represents an up-to-date, 
global indication in favor of $\theta_{13}>0$ at a confidence level of $\sim\!95\%$  ($\sim\!2\sigma$).

Figure~4 provides a final, graphical overview of the hints in Eqs.~(1)--(3) and (5).

\newpage

\section{Near-future prospects}

We have seen that several $\sim\!1\sigma$ hints seem to converge towards an overall
$\sim\!2\sigma$ indication in favor of $\theta_{13}>0$ [Eq.~(5)]. This indication
will be further tested with more accurate data in the near future. 

The SNO experiment
is completed, but the final data analysis is in progress and is being performed
with a lower energy threshold \cite{McDonald}. Updated SNO
results are expected soon. The KamLAND detector continues to take data,
but the current purification and calibration processes \cite{Suzuki} might
delay the next release of reactor neutrino data. In any case, a joint 
$3\nu$ analysis by the SNO and KamLAND collaborations would be very relevant to
to test the hint in Eq.~(1) which, if confirmed, could be possibly upgraded
to the $\sim\!\!2\sigma$ level.

Concerning atmospheric $\nu$'s, the
SK collaboration analysis is in progress
\cite{Naka}, at least for phases I+II.
It is important that all $(\delta m^2,\,\theta_{13})$-driven 
terms are included in a full $3\nu$ oscillation analysis \cite{Foglireview,Workshop}. 
Higher-statistics LBL data from the MINOS experiment 
are expected to test more accurately the possible $\nu_e$ excess \cite{Diwan}. 
Also in this case, a combination of  SK+MINOS data might then favor $\theta_{13}>0$ at
an overall $\sim\!\!2\sigma$ level, if indeed $\sin^2\theta_{13}\sim 0.02$.

In conclusion, if all the hints persists and converge in the next couple of years, 
then the current global indication in favor of $\theta_{13}>0$ [Eq.~(5)] might be promoted
to a $\sim\!\!3\sigma$ level: an exciting scenario, which would suggest early 
$\theta_{13}$ discoveries in the upcoming, next-generation experiments at reactors
 \cite{Cao,Dawson} and accelerators \cite{Mezzetto,Kajita},  
opening the door to leptonic CP violation searches.

\section{Acknowledgments}
G.L.F., E.L., A.M., and A.M.R.\ acknowledge 
support by the Italian MIUR and INFN through the ``Astroparticle Physics'' 
research project, and by the EU ILIAS through the ENTApP project. 
A.P.\  acknowledges support by MEC under the I3P program, by Spanish grants
FPA2008-00319/FPA and FPA2008-01935-E/FPA
and ILIAS/N6 Contract RII3-CT-2004-506222.
G.L.F. and E.L.\ thank Milla Baldo Ceolin for kind hospitality in Venice.

\end{document}